\documentclass{article}

\PassOptionsToPackage{numbers,compress}{natbib}

\usepackage[final]{neurips_2021}

\usepackage[utf8]{inputenc} 
\usepackage[T1]{fontenc}    
\usepackage{hyperref}       
\usepackage{url}            
\usepackage{booktabs}       
\usepackage{amsfonts}       
\usepackage{nicefrac}       
\usepackage{microtype}      
\usepackage{xcolor}         

\usepackage{graphicx}
\usepackage{hyperref}
\usepackage{amsmath,amsfonts,amssymb}
\usepackage{bm}

\DeclareMathOperator*{\argmin}{arg\,min~}
\DeclareMathOperator*{\argmax}{arg\,max~}

\newcommand{\given}{\,\vert\,}
\newcommand{\Given}{\,\Vert\,}
\DeclareMathOperator{\kl}{KL}
\DeclareMathOperator{\E}{\mathbb{E}}

\title{Posterior Temperature Optimization in Variational Inference for Inverse Problems}

\author{%
  Max-Heinrich~Laves \\
  Hamburg University of Technology \\
  \texttt{max-heinrich.laves@tuhh.de} \\
  \And
  Malte~Tölle \\
  Heidelberg University Hospital \\
  \texttt{malte.toelle@med.uni-heidelberg.de}
  \AND
  Alexander~Schlaefer \\
  Hamburg University of Technology \\
  \texttt{schlaefer@tuhh.de} \\
  \And
  Sandy~Engelhardt \\
  Heidelberg University Hospital \\
  \texttt{sandy.engelhardt@med.uni-heidelberg.de}
}

\begin{document}

\maketitle

\begin{abstract}
Bayesian methods feature useful properties for solving inverse problems, such as tomographic reconstruction.
The prior distribution introduces regularization, which helps solving the ill-posed problem and reduces overfitting.
In practice, this often results in a suboptimal posterior temperature and the full potential of the Bayesian approach is not realized.
In this paper, we optimize both the parameters of the prior distribution and the posterior temperature
using Bayesian optimization.
Well-tempered posteriors lead to better predictive performance and improved uncertainty calibration, which we demonstrate for the task of sparse-view CT reconstruction.
Our source code is publicly available at \href{https://github.com/Cardio-AI/mfvi-dip-mia}{github.com/Cardio-AI/mfvi-dip-mia}.
\end{abstract}

\section{Introduction}

Reconstructing a tomography from a finite number of X-ray projections requires solving an inverse problem.
The unknown tomography $ \bm{x} $ can only be observed through projections $ \bm{y} = \mathcal{F}[\bm{x}] $, affected by the forward Radon transform $ \mathcal{F} $, which is not directly invertible.
The reconstruction can be found by minimization of the ill-posed objective $ \hat{\bm{x}} = \argmin \{ \mathcal{L}(\bm{y}, \mathcal{F}[\hat{\bm{x}}]) + \lambda \mathcal{R}(\hat{\bm{x}}) \} $, with similarity measure $ \mathcal{L} $ and regularization $ \mathcal{R} $, weighted by $ \lambda $ \cite{Sotiras2013}.
Common regularization is manually engineered, such as penalization of spatial derivatives, or implicitly learned from a large data set.
However, obtaining ground truth pairs $ \{ \bm{x}, \bm{y} \} $ is impossible in computed tomography (CT), especially in sparse-view CT, where only a limited number of projections are obtained to reduce radiation exposure.

Deep image prior (DIP) has shown promising results in solving inverse problems by optimizing a randomly-initialized convolutional network as neural representation of the reconstruction \cite{Lempitsky2018,Baguer2020}.
To overcome the overfitting behavior of DIP, different Bayesian approaches have been proposed \cite{Cheng2019,Laves2020MCDIP}.
In Bayesian deep learning, a prior distribution $ p(\bm{w} \given \alpha) $ is placed over the weights $ \bm{w} $ of a neural network, governed by a hyperparameter $ \alpha $.
After observing the data $ \mathcal{D} $, we are interested in the posterior $ p(\bm{w} \given \mathcal{D}, \alpha) = p(\mathcal{D} \given \bm{w}, \alpha) p(\bm{w} \given \alpha) / p(\mathcal{D}) $.
However, this distribution is not tractable in general as the normalizing factor involves marginalization of the model likelihood over the prior $ p(\mathcal{D}) = \int p(\mathcal{D} \given \bm{w}, \alpha) p(\bm{w} \given \alpha) \, \mathrm{d}\bm{w} $.
A common way to approximate the posterior is variational inference (VI), which uses optimization to find the member $ q_{\bm{\phi}}(\bm{w}) $ of a family of distributions that is close to the exact posterior, defined by the variational parameters $ \bm{\phi} $.
$ q_{\bm{\phi}}(\bm{w}) $ is optimized w.r.t.\ $ \bm{\phi} $, such that the Kullback-Leibler divergence is minimized with regard to the true posterior \cite{Blei2017}.
A practical implementations of VI is \emph{Bayes by backprop}, where a fully factorized Gaussian distribution $ w_{ij} \sim \mathcal{N}(\mu_{ij}, \sigma^{2}_{ij}) $ is used as variational distribution $ q_{\bm{\phi}}(\bm{w}) $, also known as mean-field distribution, which treats the mean and variance of each weight as learnable parameters $ \phi_{ij} = \{ \mu_{ij}, \sigma^{2}_{ij} \} $ \cite{Blundell2015}.

\paragraph{Cold Posteriors}

Cold posteriors have been reported to perform better in practice in the context of Bayesian deep learning \cite{Wenzel2020}.
In order to bring the variational distribution $ q_{\bm{\phi}}(\bm{w}) $ close to the true posterior, a lower bound on the log-evidence (ELBO) is derived and maximized.
\citet{Graves2011} already suggested to reweight the complexity term in the ELBO using a factor $ \lambda $ to balance both terms in case of discrepancy between number of weights and training samples:
\begin{equation}
    \mathrm{ELBO}(q_{\bm{\phi}}(\bm{w})) = \mathbb{E}_{\bm{w} \sim q} [ \log p(\mathcal{D} \given \bm{w}) ] - \lambda \kl [q_{\bm{\phi}}(\bm{w}) \Given p(\bm{w})] ~ .
    \label{eq:elbo_lambda}
\end{equation}
It is common for Bayesian deep learning researchers to employ values of $ \lambda < 1 $ to achieve better predictive performance \cite{Blundell2015}.
While their main motivation was to qualitatively balance out discrepancies between number of model parameter and dataset size, the reweighting has recently been studied in more detail and described as the ``cold posterior'' effect \cite{Wilson2020}.
\citet{Wenzel2020} derived the tempered Bayesian posterior $ p(\bm{w} \given \mathcal{D}) \propto \exp(-U(\bm{w})/T) $ with posterior energy function $ U(\bm{w}) = - \log p(\mathcal{D} \given \bm{w} ) - \log p(\bm{w}) $ and have shown empirically that cold posteriors with $ T < 1 $ perform considerably better.
The authors also recover Eq.\,(\ref{eq:elbo_lambda}) and show that introducing $ \lambda $ into the ELBO is equivalent to a partially tempered posterior, where only the likelihood term is scaled.

In this paper, we will not argue whether cold posteriors invalidate Bayesian principles, as there is disagreement among researchers \cite{Wenzel2020,Wilson2020,Izmailov2021}, but use it in a directed way to increase predictive performance and uncertainty calibration of unsupervised sparse-view CT reconstruction with deep image prior.
This workshop paper is based on our recent journal submission \cite{Laves2021b} and extends it by additional experiments on CIFAR-10/100 (see Appendix~\ref{sec:app_cifar}).

\section{Temperature-scaled Posterior}

The ELBO for a fully temperature-scaled posterior in VI is given by (derivation in Appendix~\ref{app:elbo}):
\begin{align}
    \mathrm{ELBO}_{T}(q_{\bm{\phi}}(\bm{w})) &= - \E_{\bm{w}} \left[ \log q_{\phi}(\bm{w}) - \tfrac{1}{T} \log p(\bm{w}) \right] + \E_{\bm{w}} \left[ \tfrac{1}{T} \log p (\mathcal{D} \given \bm{w}) \right] \\
    &= - \kl \left[ q_{\bm{\phi}} (\bm{w}) \Given p(\bm{w})^{\nicefrac{1}{T}} \right] + \E_{\bm{w}} \left[ \tfrac{1}{T} \log p (\mathcal{D} \given \bm{w}) \right] ~ .
\end{align}
The KL contains the scaled prior $ p_{T}(\bm{w}) \propto p(\bm{w})^{\nicefrac{1}{T}} $, which will have the same mean, but different variance as the unscaled prior.
In case of a Gaussian prior $ p(\bm{w}) \propto \exp(-\Vert \bm{w} \Vert^{2} / 2 \sigma^{2}) $, this is equivalent to a scaled prior variance $ p(\bm{w})^{1/T} \propto \exp(-\Vert \bm{w} \Vert^{2} / 2 \sigma_{T}^{2}) $ with $ \sigma_T = \sqrt{T} \sigma $ \cite{Aitchison2021}.
Therefore, we set $ p_{T} (\bm{w}) = \mathcal{N}(\bm{0}, \frac{\sigma^{2}}{T} \bm{I}^{2}) $, which results in the following minimization criterion
\begin{equation}
    \argmin_{\bm{\phi}} T \cdot \kl \left[ q_{\bm{\phi}} (\bm{w}) \Given p_{T}(\bm{w} \given T) \right] - \E_{\bm{w}} \left[ \log p (\mathcal{D} \given \bm{w}) \right] ~ ,
    \label{eq:scaled_criterion}
\end{equation}
which, in contrast to Eq.\,(\ref{eq:elbo_lambda}) and \citet{Wenzel2020}, optimizes the fully temperature-scaled $ \mathrm{ELBO}_{T} $.

\section{Posterior Temperature Optimization}
\label{sec:bo}

Instead of manually selecting the optimal posterior temperature using heuristics or inefficient grid search, we employ Bayesian optimization (BO) to jointly find the posterior temperature $ T $ and prior scale $ \sigma $.
BO allows us to optimize functions that are expensive to evaluate, e.g., the training of a deep network \cite{Snoek2015}.
It uses a computationally inexpensive surrogate to retrieve a distribution over functions.

We apply optimization of the posterior temperature to maximize the peak signal-to-noise ratio (PSNR) between the sparse-view reconstruction $ \hat{\bm{x}} $ and the dense-view image $ \bm{x} $ as a function of $ T $ and $ \sigma $
\begin{equation}
    \max_{T \in \mathcal{T}, \sigma \in \mathcal{S}} ~ f(T, \sigma) = \max_{T \in \mathcal{T}, \sigma \in \mathcal{S}} ~  \mathrm{PSNR}(\hat{\bm{x}}(T, \sigma), \bm{x})
    \label{eq:bo_objective}
\end{equation}
using a Gaussian process (GP) as surrogate $ f \sim \mathcal{GP} $.
In each step of the BO, we evaluate our objective function $ f $ at the current candidates $ T^{\ast} $ and $ \sigma^{\ast} $ to increase the set of observations $ \mathcal{D}_{\mathrm{BO}} $ and update the posterior of the surrogate model.
Next, we maximize an acquisition function $ a(T, \sigma ; \mu_{\mathcal{GP}}, \sigma^{2}_{\mathcal{GP}}) $ using the current GP posterior mean $ \mu_{\mathcal{GP}} $ and variance $ \sigma^{2}_{\mathcal{GP}} $.
Its maximizing arguments $ T^{\ast}, \sigma^{\ast} \leftarrow \argmax a(T, \sigma ; \mu_{\mathcal{GP}}, \sigma^{2}_{\mathcal{GP}}) $ are used as candidates for the next iteration \cite{Frazier2018}.
We choose the commonly accepted expected improvement (EI) as acquisition function
\begin{equation}
    a_{\mathrm{EI}}(T, \sigma ; \mu_{\mathcal{GP}}, \sigma^{2}_{\mathcal{GP}}) = \mathbb{E} \left[ \max (y - f^{\ast}), 0) \given y \sim \mathcal{N} ( \mu_{\mathcal{GP}}(T, \sigma), \sigma^{2}_{\mathcal{GP}}(T, \sigma) ) \right] ~ ,
    \label{eq:bo_ei}
\end{equation}
where $ f^{\ast} = f(T_{\mathrm{best}}, \sigma_{\mathrm{best}}) $ is the minimal value of the objective function observed so far.
Eq.\,(\ref{eq:bo_ei}) can be solved analytically as shown in \cite{Jones1998}.
We utilize automatic differentiation from modern deep learning frameworks to optimize the acquisition function to get the next candidates $ T^{\ast} $ and $ \sigma^{\ast} $ \cite{Gardner2018}.

\section{Experiments}

\begin{figure*}
    \centering
    \parbox[t]{3.3cm}{\centering\footnotesize dense-view\vphantom{g}\\
    \includegraphics[height=3.2cm]{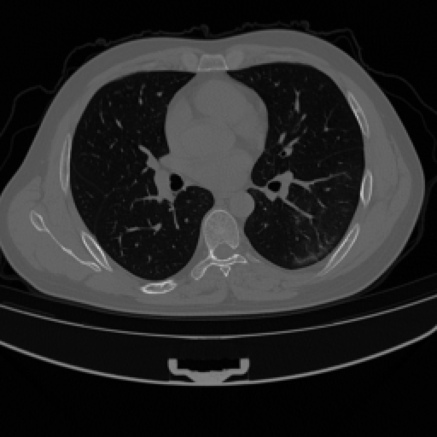}} ~ 
    \parbox[t]{3.3cm}{\centering\footnotesize FBP\vphantom{g}\\
    \includegraphics[height=3.2cm]{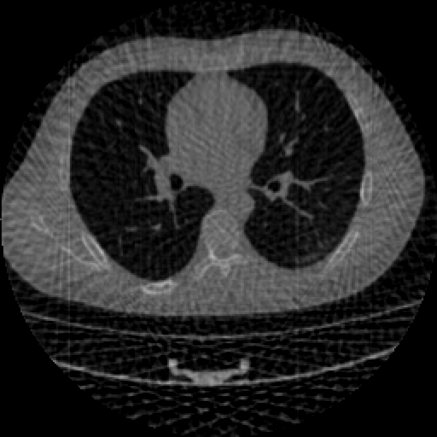}} ~ 
    \parbox[t]{3.3cm}{\centering\footnotesize DIP\vphantom{g}\\
    \includegraphics[height=3.2cm]{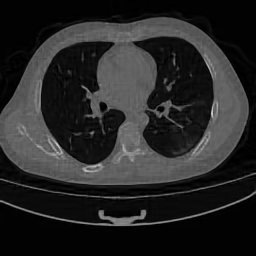}} ~ 
    \parbox[t]{3.3cm}{\centering\footnotesize $ T^{\ast}, \sigma^{\ast} $\vphantom{g}\\
    \includegraphics[height=3.2cm]{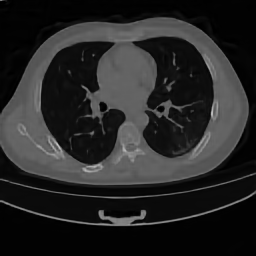}} ~ 
    \\[2mm]
    \parbox[t]{3.3cm}{\centering squared error\vphantom{g}\\
    \includegraphics[height=3.2cm]{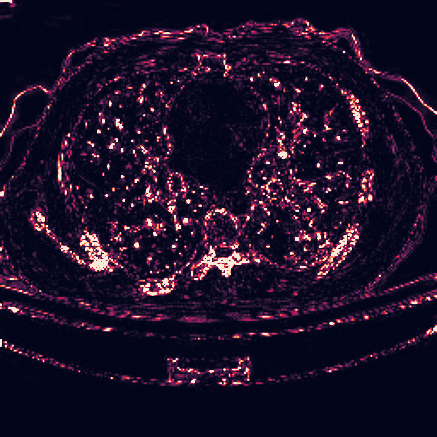}} ~ 
    \parbox[t]{3.3cm}{\centering uncertainty\vphantom{g}\\
    \includegraphics[height=3.2cm]{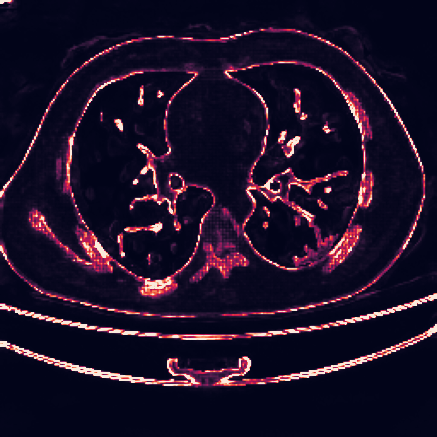}} ~ 
    \parbox[t]{3.3cm}{\centering Bayesian optimization\vphantom{g}\\
    \includegraphics[trim=2mm 0 21cm 0,clip,height=3.1cm]{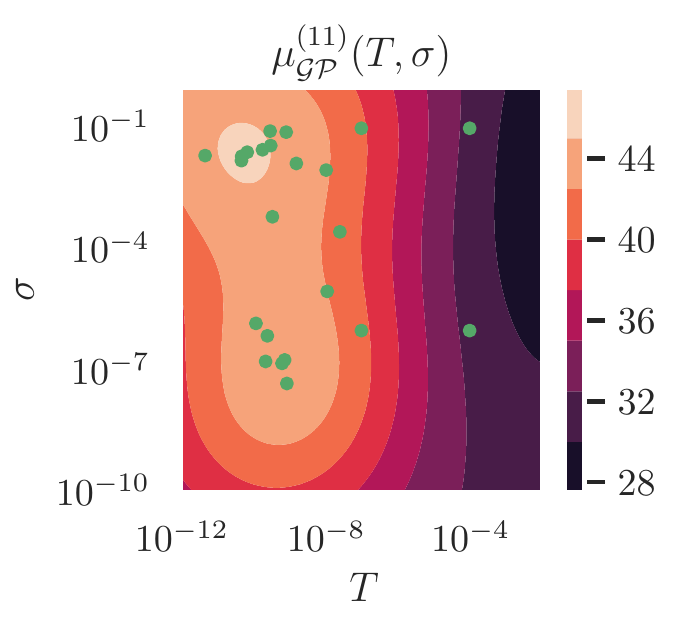}} ~
    \begin{minipage}[t][3cm][c]{3.3cm}
        \centering
        \footnotesize
        \begin{tabular}[b]{cc}
            \hline
            Method & PSNR \\
            \hline
            FBP    & 25.3 \\
            DIP    & 34.2 \\
            \hline
            $ T_1, \sigma_1 $ & 26.3 \\
            $ T_2, \sigma_2 $ & 32.8 \\
            $ T_3, \sigma_3 $ & 32.4 \\
            $ T^{\ast}, \sigma^{\ast} $ & \textbf{35.6} \\
            \hline
        \end{tabular}
    \end{minipage}
    \caption{Posterior temperature optimization for sparse-view CT reconstruction: (Top) Dense-view ground truth and sparse-view test reconstruction from FBP, non-Bayesian DIP and Bayesian DIP at optimal posterior temperature $ T^{\ast} $ and prior scale $ \sigma^{\ast} $. (Bottom) Predictive error and uncertainty at $ \{ T^{\ast}, \sigma^{\ast} \} $, mean of the GP from BO, and PSNR for different methods and values of $ \{ T, \sigma \} $.}
    \label{fig:results}
\end{figure*}

To evaluate posterior temperature optimization in Bayesian inversion, we simulate sparse-view CT by computing only 45 projections from dense-view lung CTs of COVID-19 patients\footnote{We use publicly available data from \href{https://coronacases.org}{https://coronacases.org}} using the forward Radon transform.
We use mean-field VI (MFVI) as Bayesian approach to DIP for solving the inverse task (see Fig.~\ref{fig:concept} in the appendix).
The Bayesian network is used as parameterization of the reconstruction $ \hat{\bm{x}} $ and its variational parameters are optimized by minimizing Eq.~(\ref{eq:scaled_criterion}) using the squared error $ \Vert \mathcal{F}[\hat{\bm{x}}] - \bm{y} \Vert^{2} $
as likelihood.
BO is used to find optimal values for $ \{ T, \sigma \} $ as described below.

\paragraph{Finding the Optimal Posterior Temperature}

The Gaussian process regressor from §~\ref{sec:bo} is implemented in GPyTorch \cite{Gardner2018} using a constant mean function with prior $ \mathcal{N}(15, 4^{2}) $, a scaled radial basis function kernel as covariance function and a prior length-scale $ \ell = 0.3 $.
The surrogate model is trained on observations $ \{ (\log T_{i}, \log \sigma_{i}), \mathrm{PSNR}(\hat{\bm{x}}_{T_{i}, \sigma_{i}}, \bm{x}) \} $ to impose a non-negativity constraint on $ T $ and $ \sigma $.
A Gaussian likelihood with a homoscedastic noise model with prior $ \Gamma (0.1, 100) $ is used.
We limit the search space to $ T \in [1\mathrm{e}{-12}, 1\mathrm{e}{-2}] $ and $ \sigma \in [1\mathrm{e}{-10}, 1] $ and initialize the BO with four candidate pairs with $ T \in \{ 1\mathrm{e}{-7}, 1\mathrm{e}{-4} \} $ and $ \sigma \in \{ 1\mathrm{e}{-6}, 1\mathrm{e}{-1} \} $.
If the acquisition function from Eq.\,(\ref{eq:bo_ei}) has multiple local maxima, we select the best four candidates for the next iteration.

\paragraph{Results}

The results for a test image are summarized in Fig.~\ref{fig:results}.
At optimal temperature $ T^{\ast} $, the Bayesian reconstruction outperforms filtered back-projection (FBP) and non-Bayesian DIP by means of PSNR.
From the GP mean, we see that the posterior temperature has a considerable effect on the reconstruction, with $ T^{\ast} \ll 1 $.
The effect of the prior scale is less prominent, with optimal value $ \sigma^{\ast} \approx 1\mathrm{e}{-2} $.
We observe similar findings for classification experiments on CIFAR-10/100 with Bayesian ResNets (see Appendix~\ref{sec:app_cifar}).
The uncertainty calibration is improved at optimal temperature.

\section{Conclusion}

We optimized the ELBO for a fully tempered posterior to exploit the cold posterior effect in Bayesian deep learning.
%
For ill-posed inverse problems, the optimized posterior temperature introduces the right amount of regularization to allow enough flexibility but to avoid overfitting.
This can be used in many medical applications such as CT reconstruction, registration, denoising, or artifact removal.

\clearpage

\begin{ack}
MT is supported by Informatics for Life founded by the Klaus Tschira Foundation.
ML and AS are partially funded by the Interdisciplinary Competence Center for Interface Research (ICCIR).
\end{ack}

\small
\bibliographystyle{unsrtnat}
\bibliography{bdl}
\clearpage

\normalsize
\appendix

\section{Conceptual Overview}
\label{app:overview}

\begin{figure}[ht]
    \centering
    \includegraphics[width=\textwidth]{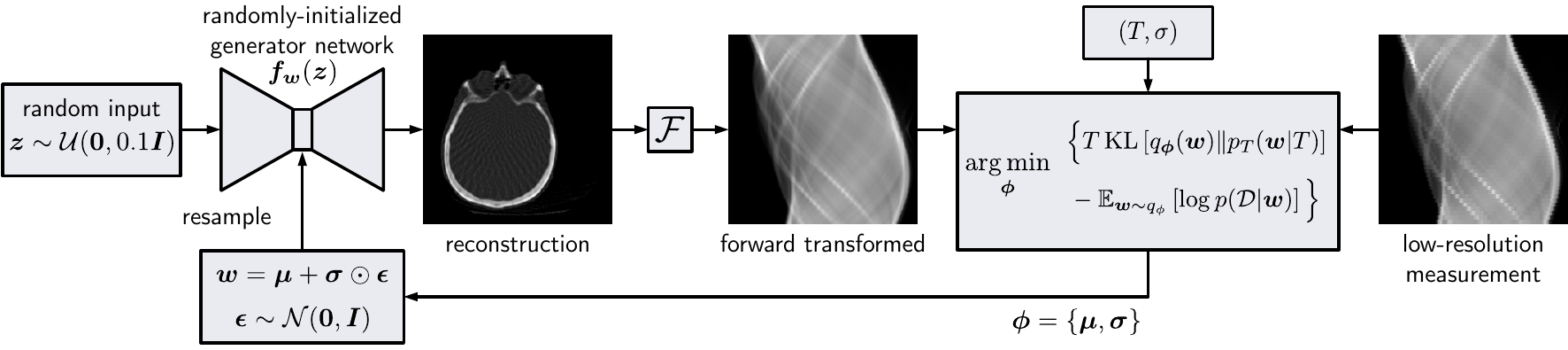}
    \caption{Conceptual overview. A randomly-initialized MFVI autoencoder network fed with uniform noise outputs a CT. The image reconstruction is performed iteratively by applying the forward Radon transform $ \mathcal{F} $ and minimizing the fully tempered negative ELBO w.r.t.\ the variational parameters $ \bm{\phi} = \{ \bm{\mu}, \bm{\sigma} \} $ using gradient descent. The posterior temperature $ T $ and prior standard deviation $ \sigma $ are found using Bayesian optimization.}
    \label{fig:concept}
\end{figure}

\section{Derivation of Fully Tempered ELBO}
\label{app:elbo}

In the following, the ELBO for a fully temperature-scaled Bayesian posterior in variational inference is derived.
Let $ p_{T}(\bm{w} \given \mathcal{D}) $ be the fully tempered posterior \cite{Wenzel2020}:
\begin{align}
    & \kl \left[ q_{\phi}(\bm{w}) \Given p_{T}(\bm{w} \given \mathcal{D}) \right] \\
    &= \E_{\bm{w}} \left[ \log q_{\phi}(\bm{w}) - \log p_{T} (\bm{w} \given \mathcal{D}) \right] \\
    &= \E_{\bm{w}} \left[ \log q_{\phi}(\bm{w}) - \log \frac{(p(\bm{w} \given \mathcal{D}) p(\bm{w}))^{1/T}}{\int (p(\bm{w}' \given \mathcal{D}) p(\bm{w}'))^{1/T} \, \mathrm{d} \bm{w}'} \right] \\
    &= \E_{\bm{w}} \left[ \log q_{\phi}(\bm{w}) - \log (p(\bm{w} \given \mathcal{D}) p(\bm{w}))^{1/T} \right] + \underbrace{\log \int (p(\bm{w} \given \mathcal{D}) p(\bm{w}))^{1/T} \, \mathrm{d} \bm{w}}_{\mathrm{const.\,w.r.t.\,} \bm{w},~ =: \log E_{T}} \\
    &= \underbrace{\E_{\bm{w}} \left[ \log q_{\phi}(\bm{w}) - \tfrac{1}{T} \log p(\bm{w}) \right] - \E_{\bm{w}} \left[ \tfrac{1}{T} \log p (\mathcal{D} \given \bm{w}) \right]}_{=: \mathrm{ELBO}_{T}(q_{\bm{\phi}}(\bm{w}))} + \log E_{T} \\
    &\Rightarrow \log E_{T} = \kl \left[ q_{\bm{\phi}}(\bm{w}) \Given p_{T}(\bm{w} \given \mathcal{D}) \right] + \mathrm{ELBO}_{T} (q_{\bm{\phi}}(\bm{w}))
\end{align}
As the tempered evidence $ E_{T} $ is constant, maximizing $ \mathrm{ELBO}_{T} $ minimizes the KL, thus bringing the variational distribution $ q_{\bm{\phi}}(\bm{w}) $ closer to the fully tempered posterior $ p_{T}(\bm{w} \given \mathcal{D}) $.

\section{CIFAR-10/100 Experiments}
\label{sec:app_cifar}

\begin{figure}[ht]
    \centering
    \includegraphics[width=1\textwidth]{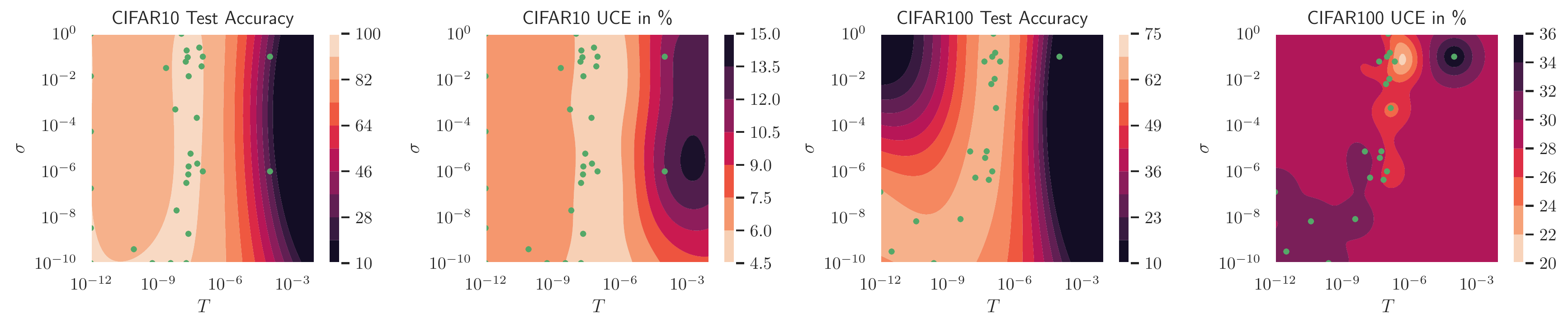}
    \caption{We additionally perform classification experiments on CIFAR-10 (ResNet-34) and CIFAR-100 (ResNet-50). The figures show estimated accuracy and uncertainty calibration error (UCE) \cite{Laves2019NIPS} landscapes. Green dots denote observed points during BO. As for CT reconstruction, the posterior temperature $ T $ has a considerable effect on both the accuracy and calibration. On CIFAR-100, the effect of the prior scale $ \sigma $ on the calibration can not be neglected. We measure uncertainty as the entropy of the softmax vector after Monte Carlo integration $ \mathcal{H} \left[ 1/N \sum^{N}_{i=1} \bm{p}(\bm{y} \given \bm{x}, \bm{w}_{i}) \right] $.}
    \label{fig:app_bo_cifar}
\end{figure}

\clearpage
\section{BO Steps}

\begin{figure}[ht]
    \centering
    \includegraphics[width=1\textwidth]{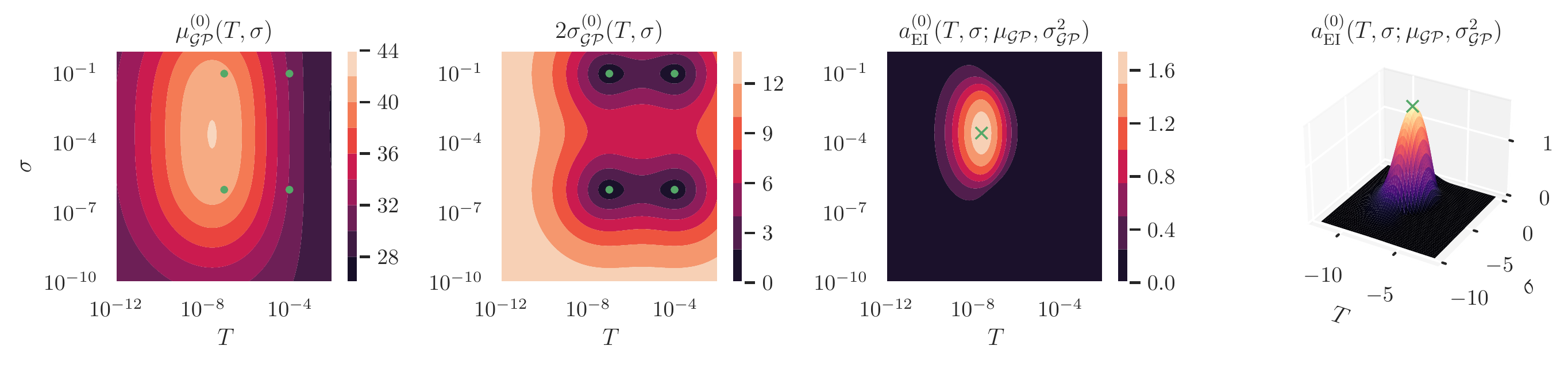}
    \includegraphics[width=1\textwidth]{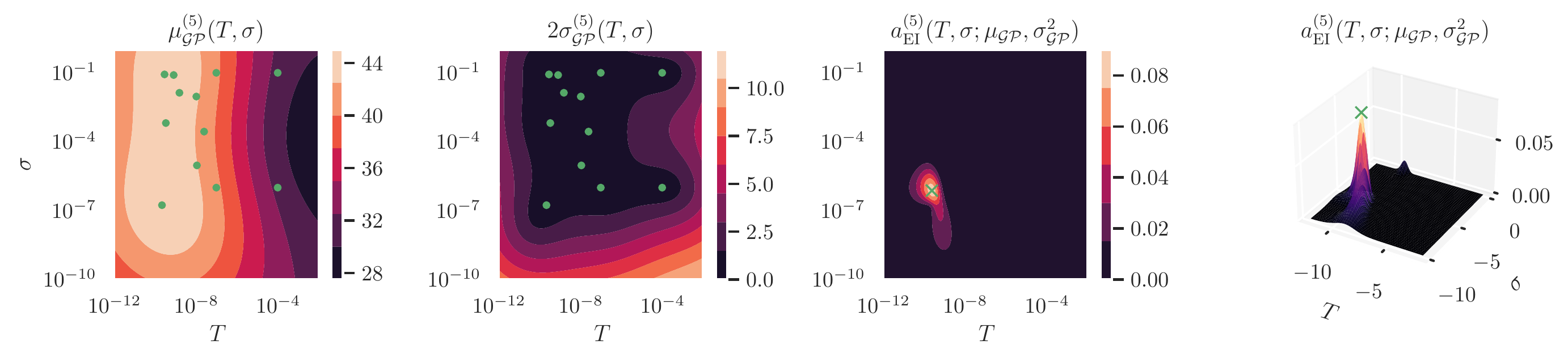}
    \includegraphics[width=1\textwidth]{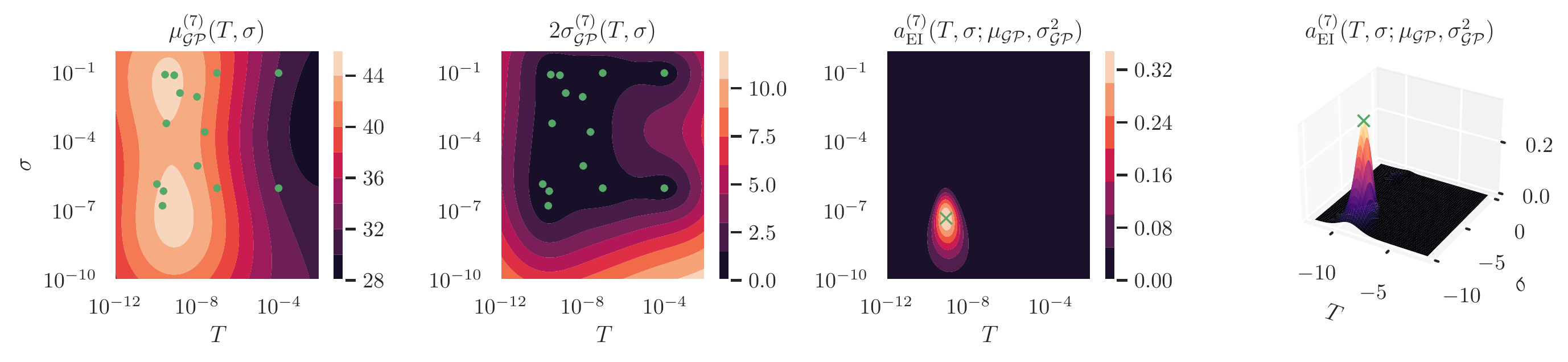}
    \includegraphics[width=1\textwidth]{img/11_fig_3d.pdf}
    \caption{Posterior temperature optimization for CT reconstruction: GP mean, confidence (2 standard deviations) and expected improvement acquisition function after BO iteration $ i \in \{0, 5, 7, 11\} $. Green dots denote observed points and green crosses show candidates for the next BO iteration. Note that per BO step, up to 4 candidates are evaluated in parallel.}
    \label{fig:app_bo_mfvi_ct}
\end{figure}

\section{Implementation Details}

\begin{itemize}
    \setlength{\itemsep}{0pt}
    \item Code for training pipeline and evaluation is available at \href{https://github.com/Cardio-AI/mfvi-dip-mia}{github.com/Cardio-AI/mfvi-dip-mia}.
    \item For CT reconstruction, we use the same architecture as described by \citet{Lempitsky2018} and optimize the network for $ 1\mathrm{e}{5} $ iterations.
    \item The final CT is sampled from the probabilistic neural representation using Monte Carlo integration $ \hat{\bm{x}} = \frac{1}{N} \sum_{i=1}^{N} \hat{\bm{x}}_{i} $, where $ \hat{\bm{x}}_{i} $ is a sample from the posterior predictive $ p(\bm{x} \given \bm{w}_{i}, \bm{y}) $.
    \item We estimate reconstruction uncertainty using the predictive variance from Monte Carlo samples $ \hat{\bm{\Sigma}} = \frac{1}{N} \sum_{i=1}^{N} (\hat{\bm{x}}_{i} - \frac{1}{N} \sum_{i=1}^{N} \hat{\bm{x}}_{i} )^{2} $.
\end{itemize}

\clearpage

\end{document}